\documentclass[10pt]{iopart}
\usepackage{epsfig}

\newcommand{\bara}{\begin{array}{c}}
\newcommand{\eara}{\end{array}}

\def\be{\begin{equation}}
\def\ee{\end{equation}}
\def\ba{\begin{eqnarray}}
\def\ea{\end{eqnarray}}
\def\bann{\begin{eqnarray*}}
\def\eann{\end{eqnarray*}}
\def\benn{\begin{displaymath}}
\def\eenn{\end{displaymath}}
\def\nn{\nonumber}
\def\lapproxeq{{\ \lower 0.6ex \hbox{$\buildrel<\over\sim$}\ }}
\def\gapproxeq{{\ \lower 0.6ex \hbox{$\buildrel>\over\sim$}\ }}
\begin{document}
\titlepage
\begin{flushright}
\vspace*{-2cm}
DTP/99/112\\
hep-ph/9912300\\
December 1999 \\
\end{flushright}

\vspace{-.2cm}

\article[Sudakov resummation for $W,\ Z$ production]{}{Sudakov logarithm resummation for vector boson
  production at hadron colliders~\footnote[3]{Contribution to the Proceedings of the UK Phenomenology Workshop on Collider Physics, Durham,
UK, 19-24 September 1999, to be published in J. Phys. G.}}
\author{A Kulesza\dag~and W J Stirling\dag\ddag}
\address{\dag\ Department of Physics, University of Durham,  
Durham DH1~3LE, U.K.}
\address{\ddag\ Department of Mathematical Sciences, University of Durham,  
Durham DH1~3LE, U.K.}

\begin{abstract}
A complete description of $W$ and $Z$ boson production at high-energy
colliders requires the resummation of large Sudakov logarithms
which dominate the production at small
transverse momentum. Currently there are two techniques for performing this
resummation: impact parameter space and  transverse momentum space. 
We argue that the latter can be formulated in a way which 
retains the advantages of the former, while at the same time allowing a 
smooth transition to finite order dominance at high transverse momentum.
\end{abstract}

\section{Introduction}

The search for a correct description of initial gluon radiation has 
recently become a very active field of theoretical research, 
stimulated by the large amount of
experimental data now available. In particular, gluon radiation plays
an important role in vector boson ($W,Z$) production at hadron colliders, 
i.e. processes that provide precision measurements of Standard Model parameters 
and appear to be significant
backgrounds to new physics phenomena. For example, one of the methods of
determining $M_W$ requires a very accurate description of the $W$'s 
transverse momentum distribution. At small transverse momentum ($q_T$), this
distribution is dominated by large logarithms $\ln(Q^2/q_T^2)$ which are directly related
to the initial emission of soft gluons. Therefore at sufficiently small $q_T$ 
fixed-order perturbation theory breaks down and the logarithms must be resummed.
 
The origin of the large logarithms is visible already at the leading--order (LO) level ---
the contribution from real emission diagrams for $q\bar{q} \rightarrow Vg$ contains a 
term of the form ${\alpha_S C_F \over \pi q_T^2}\ln\left({Q^2 \over q_T^2}\right)$.

In the case of higher-order contributions, where more gluons are emitted,
the logarithmic divergence becomes even stronger. It can be shown that in
the approximation of {\it soft and collinear} gluons with strongly ordered transverse
momenta $k_T$, i.e. 
\be
k_{T_{i1}}^2 \ll k_{T_{i2}}^2 \ll ...\ll k_{T_{iN}}^2 \lapproxeq q_T^2 \ll
Q^2
\label{strong} 
\ee
the dominant contributions to the cross section are
\ba
 {1 \over \sigma_0} {d \sigma \over d q_T^2}= &{1\over q_T^2} \Bigg[ \alpha_s {A^{(1)} 
    \over 2 \pi} \ln \left({Q^2 \over q_T^2} \right) - \alpha_s^2 {(A^{(1)})^2 
    \over 8 \pi^2 } \ln^3 \left({Q^2 \over q_T^2} \right)+... \nn \\
   & +\alpha_s^N { (-1)^{N-1} (A^{(1)})^N 
    \over 2^{2N-1}(N-1)! \pi^N} \ln^{2N-1} \left({Q^2 \over q_T^2}
    \right)+...\Bigg] \,,
\label{all_0}
\ea
where $A^{(1)}=2C_F$.
Due to the $\alpha_s^N \ln^{2N-1} (Q^2 /q_T^2)$ structure, this approximation is 
commonly known as the {\em Double Leading Logarithm 
Approximation} (DLLA). Although the approximation of 
truncating the series in (\ref{all_0}) at finite order clearly
breaks down once  $\alpha_s \ln^2 (Q^2 /q_T^2) \sim
1 $, the form of the coefficients allows (\ref{all_0}) to
be resummed, giving a so-called Sudakov factor~\cite{DDT}:  
\be
{1 \over \sigma_0} {d \sigma \over d q_T^2}= {\alpha_s A \over 2\pi q_T^2}\ln
\left({Q^2 \over q_T^2} \right) \exp \left({-\alpha_s A \over 4\pi }\ln^2
\left({Q^2 \over q_T^2} \right)\right) \,.
\label{eq_sud}
\ee

Resummation of~(\ref{all_0})  to all orders gives a finite but
unphysically {\em suppressed} result in the small $q_T$ limit. This
suppression is caused by the vanishing of strongly-ordered phase-space, in which 
overall transverse momentum conservation is ignored. The result in (\ref{eq_sud}) 
corresponds to a configuration in which a {\em  single} soft gluon balances the 
$W$ transverse momentum, giving the overall $\ln(Q^2 /q_T^2)/q_T^2$ term,
while all other gluons have  transverse  momenta $\ll q_T$. This is {\em not}
the dominant configuration in the small $q_T$ limit. Equally important are
non-strongly-ordered contributions corresponding to the emission of soft ($\sim q_T$)
 gluons whose transverse momenta add vectorially to give the overall
$q_T$ of the $W$. Although such contributions are formally subleading 
order-by-order, they do dominate the cross section in the region where
the Sudakov form factor suppresses the (formally) leading DLLA contributions.
The non-leading `kinematical' logarithms are correctly taken into account
by imposing transverse momentum conservation (rather than strong ordering),
and this is most easily achieved in Fourier transform space ($b-$space).

\section{$b-$space method }

The $b-$space method is based on introducing a $2-$dimensional impact parameter
vector $\bi{b}$,  the Fourier conjugate of $\bi q_T$~\cite{PP}. First, the
cross-section (with a delta function for transverse momentum conservation included)
is Fourier transformed, then the conjugated cross section
$\hat{\sigma}(b)$ is computed, and  finally this is transformed back to momentum space.
The advantage of the $b-$space method is that the soft gluon factorisation property is retained
even in the presence of transverse momentum conservation: 
\be
\delta^{(2)} \left( \sum^N_{i=1} \bi{k_{T_{i}}} -\bi{q_T} \right) =\int d^2 b 
{ 1 \over 4 \pi^2} e^{-i \bi {b}\bi{q_T}} 
\prod^{N}_{i=1}e^{i \bi {b}\bi{k_{T_{i}}}}\,.
\ee
This allows for the development of  a general expression resumming
all terms of the perturbation series which are at least as singular as ${1 /q_T^2}$ when
$q_T \rightarrow 0$~\cite{CSS} (at the parton level):
\be
{d\sigma \over d q_T^2} ={\sigma_0 \over 2}\int_{0}^{\infty} b d b  \, J_{0}( q_{\tiny T}
b) e^{S(b,Q^2)} \,,
\label{b_space}
\ee
with $\sigma_0 = 4 \pi \alpha^2 / (9 s)$, and  where
\ba
\label{eq:abseries}
S(b,Q^2) = - \int_{b_0^2 \over b^2}^{Q^2} \frac{d\bar\mu^2}{\bar\mu^2} 
\bigg[ \ln \bigg ( {Q^2\over\bar \mu^2} \bigg ) A(\alpha_S(\bar\mu^2)) +
B(\alpha_S(\bar\mu^2)) \bigg ] \,,\label{Sbs} \\
A(\alpha_S) = \sum^\infty_{i=1} \left(\frac{\alpha_S}{2 \pi} \right)^i A^{(i)}\
, \quad
B(\alpha_S) = \sum^\infty_{i=1} \left(\frac{\alpha_S}{2 \pi} \right)^i B^{(i)}\
,\label{AB}
\ea
with $b_0=2\exp(-\gamma_E)$.
The first two coefficients in each series~(\ref{AB}) can be
obtained~\cite{DS} from the exact fixed-order perturbative calculation 
in the high $q_T$ region by comparing
the logarithmic terms therein with the corresponding logarithms generated by
the first three terms of the expansion of $\exp(S(b,Q^2))$ in~(\ref{b_space}).

Although the $b-$space method succeeds in recovering a finite, positive result in 
the $q_T\rightarrow 0$ limit, it suffers from several drawbacks. The
first is the difficulty of matching  the resummed and fixed-order predictions. Since
the resummation is performed in the Fourier-conjugated space one loses
control over which logarithmic terms (in $q_T-$space) are taken into
account. Therefore there is no unambiguous prescription for matching; existing 
prescriptions require `unsmooth' switching from resummed to fixed-order
calculation at some value of $q_T$. Secondly, since the
integration in~(\ref{b_space}) extends from 0 to $\infty$, it is impossible to
make predictions for {\it any} $q_T$ without having a prescription for how to deal with 
the non-perturbative regime of large $b$. One prescription is to artificially prevent
$b$ from reaching large values by replacing it with a new variable $b_*$ and 
by parametrising the non-perturbative large-$b$ region in terms of a form factor. The
cross section then reads
\bann
\fl {d\sigma \over dq^2_T}
\sim   \sum_{q} e_q^2 \int_0^1 d x_A d x_B\, \delta ( x_A x_B-\frac{Q^2}{S}) 
 {1\over 2} \int_{0}^{\infty} db\;b  \, J_{0}(q_T  b)\,\\ 
\lo \!\!\!\!\!\!\!\!\!\!\!\!\!\!\!\!\!\!  \times
\, \left \{f_{q/A}(x_A,\frac{b_0}{b_*})
\,f_{\bar{q}/B}(x_B,\frac{b_0}{b_*})  F_{ab}^{NP}(Q,b,x_A,x_B)\exp{[{\cal{S}}(b_*,Q)]} 
+ \left(q \leftrightarrow \bar{q}
\right)\right\} \, .
\eann 
where, for example, the `freezing' of $b$ at $b_*$ is achieved  by 
\bann
b_*=\frac{b}{\sqrt{1+(b/b_{\rm \, lim})^2}}\; , \qquad \qquad b_* < b_{\rm \, lim}\,,
\eann
with the parameter $b_{\rm \, lim} \sim 1/\Lambda_{\rm QCD}$ separating perturbative and 
non-perturbative physics.
The detailed form of the non-perturbative function
$F_{ab}^{NP}$ remains a matter of theoretical dispute (for a review
see~\cite{ERV}). In a very simple model in which the non-perturbative contribution arises from 
a Gaussian `intrinsic' $k_T$ distribution, one would have $F \sim \exp(-\kappa b^2)$. The data 
are not inconsistent with such a form, but suggest that the parameter $\kappa$ may have some
dependence on $Q$ and $x$.

\section{$q_T-$space method}

The difficulties mentioned above could in principle be overcome if one had a
method of performing the calculations directly in transverse momentum space.
Given an insight into which logarithmic terms are resummed, it should be fairly
straightforward to perform matching with the fixed-order result. 
Moreover, the non-perturbative input
would be required in (and would affect) only the small $q_T$ region.

Three techniques have been proposed for carrying out resummation in
 $q_T-$space~\cite{EV,FNR,KS}. For a detailed discussion of the differences
between them the reader is referred to~\cite{KS2}. 
The starting point for all techniques is the general expression in impact
parameter ($b$) space for the vector boson transverse momentum
distribution in  the Drell-Yan process~\cite{CSS} at the quark level.
In the approach of \cite{KS}, the final result (in the simplest case with fixed coupling
$\alpha_S$ and retaining  only the leading coefficient $A^{(1)}$ in the $A$ and $B$ 
series of (\ref{AB})) is of the form 
\ba
\fl {1 \over \sigma_0} {d \sigma \over d q_T^2} = 
{\lambda \over q_T^2} e^{ {-\lambda \over 2} L^2}
\sum_{N=1}^{\infty} {(-2 \lambda)^{(N-1)} \over (N-1)!} 
\sum_{m=0}^{N-1} { \left( \begin{array}{c} N-1 \\ m \end{array} \right)}
L^{N-1-m}  
 \bigg[2\tau_{N+m}+ L \tau_{N+m-1}\bigg]\,.\nn \\
\label{qt_sum2}
\ea
Here $L=\ln(Q^2 /q_T^2)$, $\lambda = \alpha_S C_F /\pi$,
and the numbers $\tau_m$ are defined by~%
\footnote{The $\tau_m$ are called  $\bar{b}_m({\infty})$ in~\cite{KS}.}
\be
\tau_m \equiv \int_{0}^{\infty} d y J_{1}(y) \ln^m({y \over b_0}) \, .
\label{b_def}
\ee
The $\tau_m$  can be calculated explicitly using the generating function
\be
\fl \sum_{m=0}^{\infty} {1 \over m!} t^m \tau_m =
e^{t \gamma_E}{\Gamma \left(1 + {t\over 2}\right) \over \Gamma \left(1 - {t\over 2}\right)} =
\exp \bigg[ -2 \sum_{k=1}^{\infty} { \zeta(2k+1) \over 2k+1} {\left( t \over
2 \right) }^{2k+1} \bigg] \,,
\label{b_form}
\ee
so that e.g. $\tau_0=1,\ \tau_1=\tau_2=0,\ \tau_3=-{1\over 2}\zeta(3)$
etc. 

Notice that by setting all $\tau_m$ coefficients (except
$\tau_0$) to zero one would immediately recover the 
DLLA form (\ref{qt_sum2}). Since there are no explicit subleading
logarithms in~(\ref{b_space}), other than those related to kinematics, the
presence of the $\tau_m$ coefficients must
correspond to relaxing the strong-ordering condition. This can be checked 
explicitly by performing the `exact' ${\cal O}(\alpha_s^2)$ calculation
in transverse momentum space:
\ba
\fl \int d^2k_{T1}d^2k_{T2}\, \left[ { \ln(Q^2/k_{T1}^2) \over k_{T1}^2 } \right]_+
\left[ { \ln(Q^2/k_{T2}^2) \over k_{T2}^2}  \right]_+
\, \delta^{(2)}(\bi{k_{T1}}  + \bi{k_{T2}}   -\bi{q_T} )  \nonumber \\
\lo   =   
{\pi \over q_T^2}\;  \left( -  L^3  +  4   \zeta(3) \right) . 
\label{exact2}
\ea
Strong ordering is equivalent to replacing the $\delta-$function by
$\delta^{(2)}(\bi{k_{T1}} -\bi{q_T}) \times$ \mbox{$\theta(k_{T1}^2 - k_{T2}^2)$} 
$+ (1 \leftrightarrow 2)$. This gives only the leading $L^3$ term on the right-hand side.
The $\zeta(3)$ term  represents the first appearance
of the (kinematic) $\tau_3$ coefficient of (\ref{qt_sum2}).

\begin{figure}[h]
\begin{center}
\mbox{\epsfig{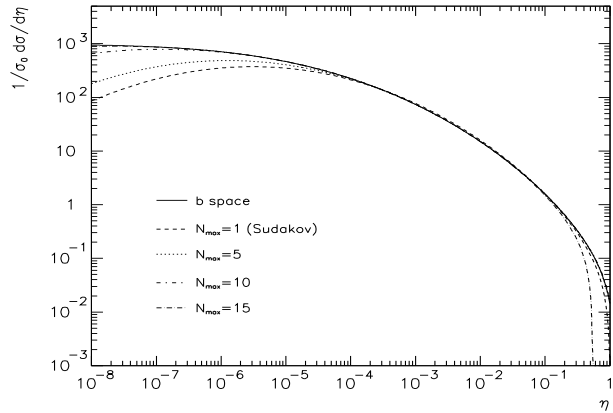}}
\end{center}
\caption{The $b$-space result (parton level, fixed coupling, only $A^{(1)}$)
  compared to the expression~(\ref{qt_sum2}), calculated for various values of
  $N_{\rm max}$. Here $\eta=q_T^2/Q^2$.} 
\label{KSvb}
\end{figure}

\begin{figure}[h]
\begin{center}
\mbox{\epsfig{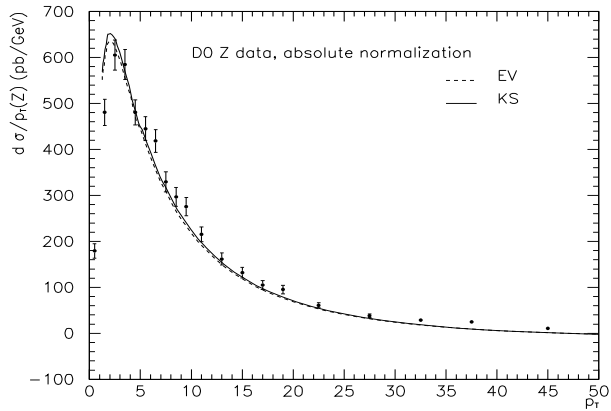}}
\end{center}
\caption{Comparison of the theoretical predictions derived in the Ellis-Veseli
  (EV)~\cite{EV} and Kulesza-Stirling (KS)~\cite{KS,KS3} approach with D0 data~\cite{D0data}.} 
\label{D0}
\end{figure}

In principle the formalism
presented above allows for an inclusion of {\em any} number of such subleading
kinematic logarithms, defined by the cut-off value $N_{\rm
  max}$. \Fref{KSvb} shows that for small values of $q_T$ the
approximation of the $b-$space result improves with  increasing $N_{\rm
  max}$. Therefore by retaining sufficiently many terms one can obtain a
  good approximation (i.e. adequate for phenomenological purposes) 
  to the $b-$space result by summing logarithms directly in $q_T$ 
space.\,\footnote{However, due to the lack of knowledge of $A^{(3)},\;B^{(3)}$, etc. it
  is only possible to obtain a result where not more than the first four
  `towers' of logarithms are fully resummed, see \cite{KS}.} 
The technique developed so far can be extended to include subleading $A$ and
$B$ coefficients, the running coupling and parton distributions, thus
yielding a `realistic' expression for the hadron-level cross section. The
result is too lengthy to reproduce here, but can be found in \cite{KS, KS3}. 
A comparison between the (Tevatron) D0 data \cite{D0data} on the $Z$ transverse momentum distribution
and two of the $q_T-$space predictions \cite{EV,KS} is
shown in~\Fref{D0}. No non-perturbative contribution is included, and in fact
the fits can be improved slightly by including a modest amount of Gaussian
intrinsic $k_T$ smearing, as discussed above.

\section{Conclusions}

We have discussed two methods of performing resummation
in vector boson production at hadron colliders. We have argued in favour of
the $q_T-$space method, which can be extended to systematically take into account
subleading logarithms of kinematic origin. However more work is required. 
 Although the $q_T-$space
method provides a simple matching prescription, the form of the non-perturbative
function in this approach remains an open theoretical issue. In particular,
the current lack of understanding of the  $x$ and $Q$ dependence of the non-perturbative
contribution is a serious limiting factor in predicting the $q_T \to 0 $ behaviour 
of the distribution at the LHC.


\end{document}